\documentclass{ws-procs9x6}
\def\Msun{\hbox{$\rm\thinspace M_{\odot}$}}

\newcommand{\ltsim}{\mbox{{\raisebox{-0.4ex}{$\stackrel{<}{{\scriptstyle\sim}}
$}}}}

\begin{document}

\title{On the cosmological evolution of quasar black-hole masses}

\author{R.~J. McLure and J.S. Dunlop}

\address{Institute for Astronomy,\\
University of Edinburgh,\\ 
Royal Observatory,\\
Edinburgh, EH9 3HJ, UK\\
E-mail: rjm@roe.ac.uk}


\maketitle

\abstracts{
Virial black-hole mass estimates are presented for 12698 quasars
in the redshift interval $0.1\leq z \leq 2.1$, based on modelling of 
spectra from the Sloan Digital Sky Survey (SDSS) first data release . 
The black-hole masses of the SDSS quasars are found to lie 
between $\simeq10^{7}\Msun$ and an upper
limit of $\simeq 3\times 10^{9}\Msun$, entirely consistent with the 
largest black-hole masses found to date in the local
Universe. The estimated Eddington ratios of the broad-line quasars (FWHM\,$\geq
2000$ km s$^{-1}$) show a clear upper boundary at 
$L_{bol}/L_{Edd}\simeq 1$, suggesting that the Eddington luminosity is
still a relevant physical limit to the accretion rate of 
luminous broad-line quasars at $z\leq 2$. By combining the 
black-hole mass distribution of the SDSS quasars with the 
2dF quasar luminosity function, the number density of 
active black holes at $z\simeq 2$ is estimated as a function of mass. 
By comparing the estimated number density of active black holes 
at $z\simeq 2$ with the local mass density of dormant black holes, 
we set lower limits on the quasar lifetimes and find that the 
majority of black holes with mass $\geq 10^{8.5}\Msun$ are in place 
by $\simeq 2$.}
\section{Introduction}
\begin{figure}[ht]
\centerline{\epsfxsize=4.7in\epsfbox{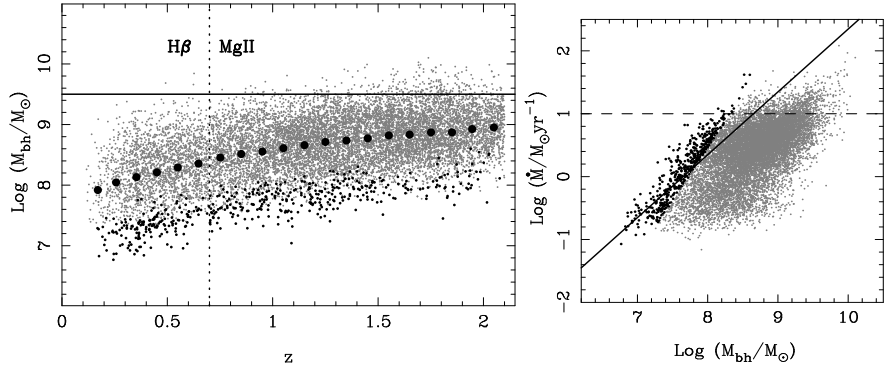}}  
\caption{
Panel A shows virial black-hole mass estimate
versus redshift for our full SDSS quasar sample. Broad-line quasars
(FWHM$\,\geq 2000$ km s$^{-1}$) are grey symbols, while
narrow-line objects (FWHM$<2000$ km s$^{-1}$) are black symbols. 
The mean black-hole masses within $\Delta z=0.1$ bins are shown as 
filled circles. The vertical dotted line highlights 
the switch from using the $H\beta$-based to the MgII-based virial 
mass estimator at $z=0.7$. The horizontal solid line marks a 
black-hole mass of $3\times 10^{9}\Msun$, the maximum mass observed at low
redshift. Panel B shows estimated accretion rate versus black-hole
mass. The solid line delineates the Eddington limit. Symbols as panel A. 
}
\end{figure}

In this proceedings we present a summary of the main results of McLure
\& Dunlop (2004) which investigates the black-hole masses of a sample
of 12698 quasars drawn from the Sloan Digital Sky Survey (SDSS) first
data release. In McLure \& Dunlop (2004) the black-hole masses 
of the SDSS quasars are estimated using the so-called virial method. 
The basic assumption
underlying this technique is that the motions of gas in the broad-line
region (BLR) of quasars are virialized. Under this assumption the mass 
of the central black-hole can be estimated 
from: $M_{bh}=G^{-1}R_{BLR}V_{BLR}^{2}$; where $R_{BLR}$ is the BLR 
radius and $V_{BLR}$ is the orbital velocity of the line-emitting
gas. The standard application of this technique uses the FWHM of the 
H$\beta$ emission line to estimate $V_{BLR}$ and the monochromatic
$5100$\AA\, luminosity to estimate $R_{BLR}$ (Kaspi et al. 2000). In
addition to this method, the analysis in McLure \& Dunlop (2004)
employs a re-calibration of the virial mass estimator in terms of the
FWHM of the MgII emission line and the $3000$\AA\, luminosity (McLure \&
Jarvis 2002).

The original sample was drawn from the SDSS quasar catalog
II (Schneider et al. 2003) which consists of some 17,000 quasars in
the redshift range $0.08~<~z~<~5.41$. For the purposes of this study
the sample was restricted to $\sim$14,000 quasars with $z<2.1$, where the upper
redshift limit is imposed by the MgII emission line being redshifted
out of the SDSS spectra. The flux-calibrated
spectra of each quasar was analysed using an automated algorithm to
consistently determined the required parameters (principally FWHMs and
continuum luminosities). After the removal of objects affected by 
low signal-to-noise, or artifacts in their spectra, the final sample 
consisted of 12698 quasars. Although not complete this sample is 
clearly representative of optically luminous quasars in the 
redshift interval $0.1<z<2.1$.

\section{Evolution of black-hole mass}

Panel A of Fig. 1 shows the virial black-hole mass estimates versus redshift
for the full sample. Two features of this figure are worthy of
individual comment. Firstly, although it can be seen that the mean black-hole
mass increases with redshift it should be remembered that, because the
mean FWHM remains approximately constant with redshift, this is
exactly as expected given the flux-limited nature of the
SDSS. Secondly, it can be seen that the distribution of the SDSS
quasar black-hole masses is entirely consistent with an upper limit 
of $\sim 3\times 10^{9}\Msun$. This limit is consistent with both 
the most massive black-holes measured dynamically in the local 
Universe, and the expected black-hole mass limit based on the 
known properties of early-type galaxies and the locally observed correlation 
between bulge and black-hole mass (i.e. $\sigma\,\ltsim\, 
400$~km s$^{-1}$ and $L\,\ltsim \,10\, L^{\star}$).  Consequently, 
in contrast to Netzer (2002), using the MgII-based virial mass 
estimator no evidence is found for a conflict between quasar
black-hole masses at $z<2.1$ and the 
contemporary, or ultimate properties of their host-galaxy population.

\section{Quasar accretion rates}

Panel B of Fig 1. shows the estimated quasar accretion rates versus black-hole
mass for the full sample, where the accretion rates have been
calculated from the estimated bolometric luminosities assuming 
a canonical mass-to-energy conversion efficiency of $\epsilon=0.1$. It
can be seen that the vast majority of the broad-line quasars (95\%)
are accreting at sub-Eddington rates. However, perhaps most
interestingly, Panel B shows there to be a significant 
absence of quasars with black-hole masses $M_{bh}\geq10^{9}\Msun$ 
accreting close to the Eddington limit. This result is consistent 
with the existence of a physical limit to the amount of gas which 
can be supplied to the central regions of quasars of $\simeq 10\Msun$yr$^{-1}$.

\section{The number density of black-holes at z=2}
\begin{figure}[ht]
\centerline{\epsfxsize=4.7in\epsfbox{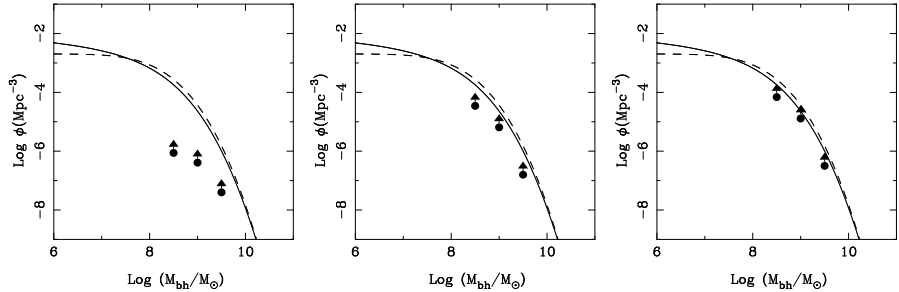}}  
\caption{In each panel the solid and dashed lines show the cumulative
local dormant black-hole mass functions as derived from 
the $M_{bh}-L_{bulge}$ and $M_{bh}-\sigma$ relations respectively
(McLure \& Dunlop 2004). In panel A we show the estimated 
number densities of active 
supermassive black-holes at $z\simeq 2$ for three mass thresholds
($\geq10^{8}\Msun, \geq10^{9}\Msun\, \& \geq10^{9.5}\Msun$). In 
panel B these number densities have be adjusted to account 
for a possible relationship between mean quasar lifetime 
and black-hole mass (see text for discussion). In Panel C the 
number densities of supermassive black-holes have been 
boosted by a factor of two to conservatively account for geometric
obscuration.}
\end{figure}

Using a sample of 372 quasars common to the SDSS DR1 and the 10K
release of the 2dF quasar catalog (Croom et al. 2001) it was possible
to accurately convert the SDSS luminosities into the absolute $B-$band
magnitudes used in the derivation of the 2dF QSO luminosity function
by Boyle et al. (2000). Using the 2dF QSO luminosity function to
calculate the number densities of quasars brighter than a given
absolute magnitude, it is then possible to use the SDSS black-hole
masses to estimate the number density of active black-holes at
a given redshift as a function of mass. The results of this 
calculation are shown in panel A of Fig 2. for a sub-sample of the 
SDSS quasars centred on $z=2$. Comparison with the local dormant 
black-hole mass function (solid/dashed lines) shows an apparently 
increasing black-hole activation fraction of $f\simeq 0.005$ at 
$M_{bh}\simeq 10^{8.5}\Msun$ rising to $f\simeq 0.05$ 
at $M_{bh}\simeq 10^{9.5}\Msun$. In panel B of Fig 2. the number
density of all black-holes in place by $z=2$ has been estimated by
correcting the number densities of panel A using the increasing
relationship between quasar lifetime and black-hole mass predicted by
Yu \& Tremaine (2002). Finally, it can be seen from  panel C of Fig 2. that,
after a conservative correction of a factor of two to account 
for geometric obscuration, the direct implication of these results is that the 
fraction of black-holes with mass $\geq 10^{8.5}\Msun$ which are in
place at $z\simeq 2$ is $\geq 0.4$.

\end{document}